\documentclass[onecolumn]{revtex4}
\usepackage{amsfonts}
\usepackage{amsmath}
\usepackage{amssymb}
\usepackage{graphicx}

\begin{document}

\title{A general formula for the amplitude-frequency ratio in shaking induced Mott insulator of atomtronic transistors}
\author{Wenxi Lai}\email{wxlai@pku.edu.cn}\author{Yu-Quan Ma}\author{Yi-Wen Wei}

\affiliation{School of Applied Science, Beijing Information Science and Technology University, Beijing, 100192, China}

\begin{abstract}
\textbf{Abstract}: Mott insulator of atomic transport can be realized in shaken optical lattices by choosing particular ratio of driving amplitude and frequency, which has been studied as Floquet engineering with time-independent effective Hamiltonian approach. Here, we give a general formula of amplitude-frequency ratio for realization of the shaking induced insulator-conductor transition in a double-well open system, using numerical computation with instantaneous eigenstates approach. The result is owing to the fact that the instantaneous eigenstates approach is applicable in wider parameter range compared with the time-independent effective Hamiltonian approach. Analysis from the results of quantum master equation shows that the insulator effect is originated from coherent localization of atom wave packets in optical wells.

\end{abstract}

\maketitle
\begin{flushleft}
  \item[1.] \textbf{Introduction}
\end{flushleft}
Achieving the full understanding and control of the insulator-conductor transition in optical lattices is significant for the future generation of atomtronic devices which focuses on atom analogs of electronic materials, devices and circuits~\cite{Seaman,Pepino,Caliga,Caliga17,Lai,JGLee,Wilsmann}. Recently, periodically shaken (or driven) optical lattices attract growing attentions, which show fine qualities in coherent control of ultracold atomic gases, for examples, coherent destruction of tunneling (CDT)~\cite{Grossmann,Grifoni,Kierig,Lignier}, transition between superfluid and Mott-insulator~\cite{Zenesini,Eckardt,Creffield6,Liberto}, fractional quantum Hall effect~\cite{Sorensen,Miao}, topological non-trivial states~\cite{Kitagawa,Rudner,Zheng,Wintersperger,Cheng,JYZhang}, topological charge pumping~\cite{Mei,Kang}, topological superradiance~\cite{Feng}, discontinuous quantum phase transitions~\cite{Song}, artificial gauge fields~\cite{Struck,Hauke,Creffield,Price}, atomic gas solitons~\cite{Blanco,Wang}, and atomic analogue of photocurrent in optical lattices~\cite{Heinze}.

Previously in shaken optical lattices, time-independent effective Hamiltonian approach plays important role for dealing with periodic time-dependent Hamiltonians~\cite{Maricq,Grozdanov,Goldman,Eckardt2017,Sun,Bastidas,Abanin,Yang2018,Jangjan,Lindner,Fleury}. Furthermore, efficiency of this approach has been proved in many experiments~\cite{Lignier,Zenesini,Rodriguez,Redondo,Maczewsky,Liu2024}. The effective Hamiltonian is generally obtained with expansion in perturbation methods~\cite{Goldman,Eckardt2017,Itin2015}. In the expansion, driving frequency $\omega$ is required to be larger than characteristic energy $E$ of corresponding systems, $\omega>E$, for good approximations~\cite{Goldman,Eckardt2017,Sun}. For lower driving frequency, one may need to consider higher order expansions in effective Hamiltonians~\cite{Creffield6}, called multi-photon processes. When driving frequency is lower than the characteristic energy $\omega<E$, Bessel-function based effective Hamiltonian approach would deviate from experimental results~\cite{Lignier}. Nonlinear corrections to the effective Hamiltonian from atom interactions were analyzed in Ref.~\cite{Itin2008} and corrections to it due to a trap were studied in Ref.~\cite{Itin}.

\begin{figure}
  \includegraphics[width=7.5cm]{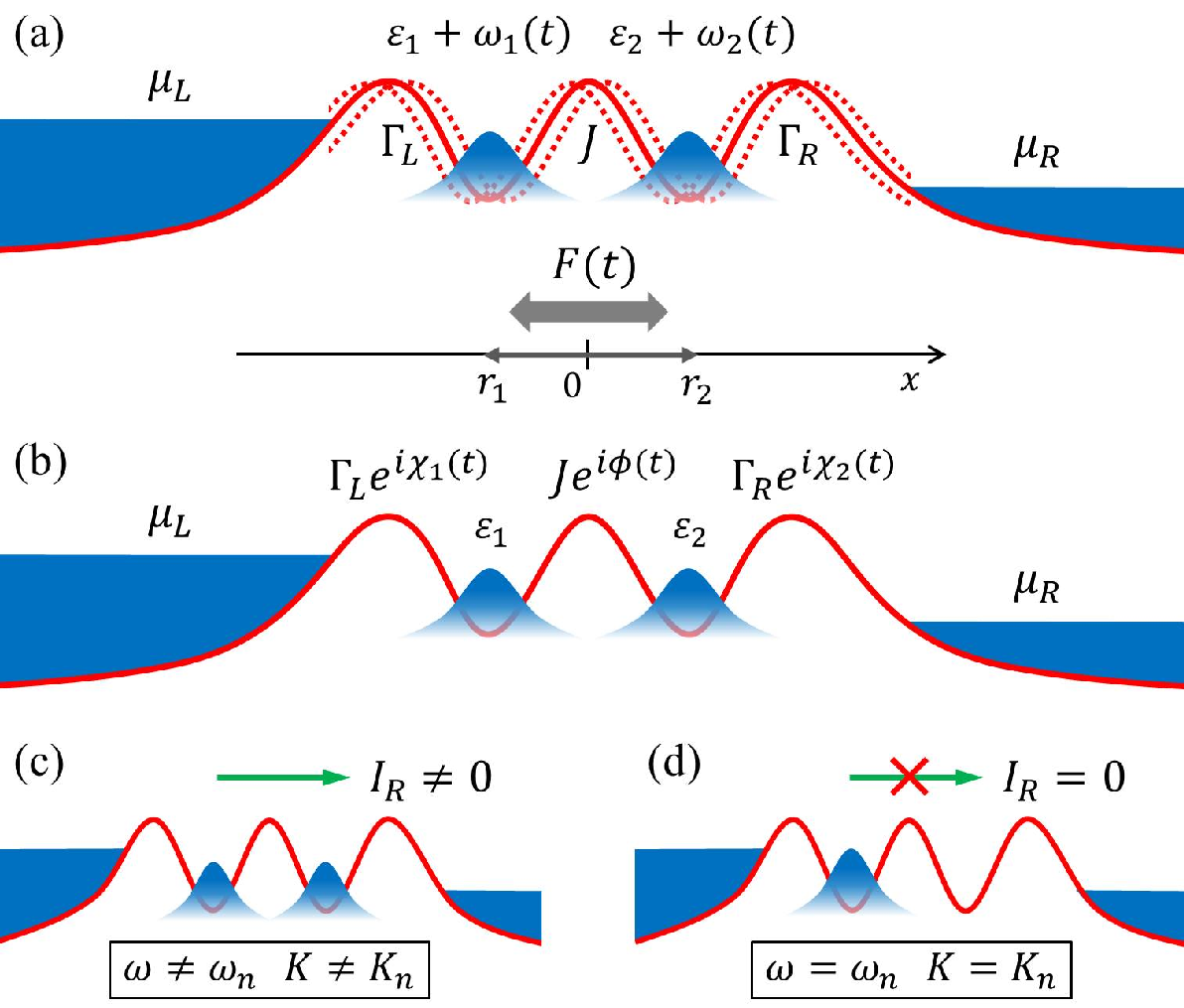}\\
  \caption{(a) In the laboratory frame, the double-well potential is shaking, coupling to two atomic baths. (b) In the oscillating frame, tunneling coefficients become time-dependent. (c) Atom wave packet occupies both the two optical wells, which leads to stationary current. (d) Atom population is trapped in the first well by the choice of driven frequency $\omega_{n}$ and driven amplitude $K_{n}$ ($n=1$, $2$, $3$,...). As a result, atomic current is stopped.}\label{fig1}
\end{figure}

Motivation of the present work is to study general condition of shaking induced insulator-conductor transition in an optical lattice transistor for the development of atomtronics. To this end, we consider a periodically driven double-well open system in optical potentials to represent the shaking atomtronic transistor. Non-equilibrium ultracold bosonic atoms are supposed to flow in the transistor under chemical potential. Actually, periodically driven open systems have been studied earlier~\cite{Grifoni,Shirai,Liu2015,Restrepo,Dai,Schnell}, using time-independent effective Hamiltonian approach mentioned above. In contrast, we directly solve the shaking open system numerically based on its time-dependent Hamiltonian, expecting this method offers more general result of the system properties comparing with the results given by the time-independent effective Hamiltonian approach. Using the instantaneous eigenstates approach, firstly, we develop a variable coefficient quantum master equation as a general approach to study driven open systems. Secondly, coherent trapping induced Mott insulator at arbitrary shaking frequency, both in $\omega\leq E$ and $\omega>E$, is demonstrated. Thirdly, for the first time, a common formula of the amplitude-frequency ratio for realization of the Mott insulator would be given here.

\begin{flushleft}
  \item[2.] \textbf{The system}
\end{flushleft}
The double-well open system in optical potentials is schematically illustrated in Fig.~\ref{fig1}. It is connected to the left and right side atomic reservoirs, respectively. Due to the chemical potential difference between the two reservoirs, individual atoms pass through the double-well system. Energy levels in the double-well represent conduction band of atomtronic circuit~\cite{Seaman,Pepino}, in which interactions between bosonic atoms are supposed to be strong enough that the maximum occupation of each level is limited to be a single atom~\cite{Qian,Krinner,MEKim,Stuart}. The model of single bosonic atom occupation results from on-site interaction between atoms because of contact interactions and dipole-dipole interactions~\cite{Wilsmann,Greiner,Gajdacz,Cao}. Atoms in the system are supposed to be suffered from shaking of optical potential. Motion of these atoms could be described by the time-dependent Hamiltonian in the laboratory reference frame
\begin{eqnarray}\label{eq:OH}
\mathcal{H}(t)&=&\mathcal{H}_{S}(t)+\mathcal{H}_{E}+\mathcal{H}_{C},
\end{eqnarray}
in which the system Hamiltonian is a time-dependent operator written by ($\hbar=1$)
\begin{eqnarray}\label{eq:HS}
\mathcal{H}_{S}(t) &=& \sum_{l=1}^{2}(\varepsilon_{l}+\omega_{l}(t))n_{l}-J (a_{1}^{\dag}a_{2}+a_{2}^{\dag}a_{1}),
\end{eqnarray}
where $a_{l}$ and $n_{l}=a_{l}^{\dag}a_{l}$ are annihilation and number operators of atoms, respectively, at lattice site $l$ ($l=1$, $2$). $\varepsilon_{l}$ denotes corresponding bare energy level. When the optical potential oscillates periodically, in the oscillating reference frame, the energy level becomes $\varepsilon_{l}+\omega_{l}(t)$ with respect to the shaking induced potential $\omega_{l}(t)=-r_{l}\cdot F(t)$, in which the inertial force is given by $F(t)=F_{\omega}\cos(\omega t)e_{x}$ with $\omega=2\pi/T$. Here, positions of the two optical wells are denoted by $r_{1}=-\frac{d}{2}e_{x}$ and $r_{2}=\frac{d}{2}e_{x}$. The two optical wells are coherently coupled with the tunneling strength $J$. The atomic electrodes are regarded as free atomic gas with the bare energy,
\begin{eqnarray}\label{eq:HB}
\mathcal{H}_{E} &=& \sum_{\alpha,k}\epsilon_{\alpha k}n_{\alpha k},
\end{eqnarray}
where $a_{\alpha k}$ and $n_{\alpha k}=a_{\alpha k}^{\dag}a_{\alpha k}$ are annihilation and number operators in the left ($\alpha=L$) and right ($\alpha=R$) electrodes, respectively. Here, $\epsilon_{\alpha k}$ represents energy of a single atom with momentum $k$. Coupling between the shaken system and atomic electrodes are described by the Hamiltonian,
\begin{eqnarray}\label{eq:HE}
\mathcal{H}_{C} &=& -\sum_{k}(t_{L}a_{1}^{\dag}a_{L k}+t_{R}a_{2}^{\dag}a_{R k}+H.c.),
\end{eqnarray}
where $t_{\alpha}$ indicates a coupling amplitude with respect to the atomic bath $\alpha$.

\begin{flushleft}
  \item[3.] \textbf{Instantaneous eigenstates approach}
\end{flushleft}

The instantaneous eigenstates approach is a method to describe dynamics of a quantum system with time dependent Hamiltonian. On one side, equation of motion of the system should be derived directly using time dependent Hamiltonian. On the other side, to solve results of the equation of motion, time dependent eigenstates and eigenvalues would be considered.

\begin{flushleft}
  \item[3.1] \textbf{Equation of motion}
\end{flushleft}
The total quantum states are represented by the density matrix $\varrho_{tot}$ which describes both the double-well system and the atomic baths. Time evolution of the density matrix $\varrho_{tot}$ satisfies the Liouville-von Neumann equation,
\begin{eqnarray}\label{eq:OEOM}
\frac{\partial}{\partial t}\varrho_{tot}(t) &=& -i[\mathcal{H}(t),\varrho_{tot}(t)].
\end{eqnarray}
It is convenient to work in the oscillating frame according to the Floquet theory~\cite{Eckardt2017,Vega}. With the gauge transformation operator $U(t)=\exp[i\sum_{l=1}^{2}\chi_{l}(t)n_{l}]$, where $\chi_{l}(t)=-\int_{0}^{t}dt'\omega_{l}(t')$, Eq. \eqref{eq:OEOM} is transformed into the oscillating frame
\begin{eqnarray}\label{eq:EOF}
\frac{\partial}{\partial t}\rho_{tot}(t) = -i[H(t),\rho_{tot}(t)].
\end{eqnarray}
It can be readily seen that the total density matrix $\rho_{tot}(t)=U^{\dag}(t)\varrho_{tot}(t)U(t)$ is governed by a time-dependent Floquet Hamiltonian
\begin{eqnarray}\label{eq:TDFH}
H(t)=U^{\dag}(t)\mathcal{H}(t)U(t)-i U^{\dag}(t)\frac{\partial}{\partial t}U(t)
\end{eqnarray}
which has the following detail expression,
\begin{eqnarray}\label{eq:TH}
H(t) &=& \sum_{l=1}^{2}\varepsilon_{l}n_{l}-J (e^{i\phi(t)}a_{1}^{\dag}a_{2}+H.c.)+\sum_{\alpha,k}\epsilon_{\alpha k}n_{\alpha k}-\sum_{k}(t_{L}(t)a_{1}^{\dag}a_{L k}+t_{R}(t)a_{2}^{\dag}a_{R k}+H.c.).
\end{eqnarray}
Now, the tunneling strength is time-dependent with the Peierls phase $\phi(t)=\chi_{2}(t)-\chi_{1}(t)$, where we have $\chi_{2}(t)=-\chi_{1}(t)=\frac{K}{2\omega}\sin(\omega t)$ with the driving strength (amplitude) $K=F_{\omega}d$. In addition, the new system-bath couplings are expressed as $t_{L}(t)=t_{L}e^{-i\chi_{1}(t)}$ and $t_{R}(t)=t_{R}e^{-i\chi_{2}(t)}$.

Next, to derive quantum master equation of the system, we separate the Hamiltonian \eqref{eq:TH} into three parts as $H(t)=H_{f}+H_{J}(t)+H_{C}(t)$, which describes free evolution $H_{f}= \sum_{l=1}^{2}\varepsilon_{l}n_{l}+\sum_{\alpha,k}\epsilon_{\alpha k}n_{\alpha k}$, the coherent tunneling $H_{J}(t)=-\sum_{\langle l,j\rangle} Je^{i\phi_{lj}(t)}a_{l}^{\dag}a_{j}$ and the system-electrode couplings $H_{C}(t)=-\sum_{k}(t_{L}(t)a_{1}^{\dag}a_{L k}+t_{R}(t)a_{2}^{\dag}a_{R k}+H.c.)$, respectively. Using unitary operator $e^{-itH_{f}}$, Eq. \eqref{eq:EOF} could be transformed into the form in interaction picture,
\begin{eqnarray}\label{eq:IE}
\frac{\partial}{\partial t}\hat{\rho}_{tot}(t)=-i[\hat{H}_{J}(t),\hat{\rho}_{tot}(t)]-i[\hat{H}_{C}(t),\hat{\rho}_{tot}(t)],
\end{eqnarray}
where all terms in the equation have been transformed into their correspondences in interaction picture, $\hat{\rho}_{tot}(t)=e^{itH_{f}}\rho_{tot}(t)e^{-itH_{f}}$, $\hat{H}_{J}(t)=e^{itH_{f}}H_{J}(t)e^{-itH_{f}}$ and $\hat{H}_{C}(t)=e^{itH_{f}}H_{C}(t)e^{-itH_{f}}$. By performing an integral on time to Eq. \eqref{eq:IE}, one reaches the equation,
\begin{eqnarray}\label{eq:IIE}
\hat{\rho}_{tot}(t)&=&\hat{\rho}_{tot}(0)-i\int_{0}^{t}dt'[\hat{H}_{J}(t'),\hat{\rho}_{tot}(t')]-i\int_{0}^{t}dt'[\hat{H}_{C}(t'),\hat{\rho}_{tot}(t')].
\end{eqnarray}
Substitution of Eq. \eqref{eq:IIE} into the last term of Eq. \eqref{eq:IE} gives rise to the second order extension with respect to the system-environment coupling,
\begin{eqnarray}\label{eq:SIIE}
\frac{\partial}{\partial t}\hat{\rho}_{tot}(t)&=&-i[\hat{H}_{J}(t),\hat{\rho}_{tot}(t)]-i[\hat{H}_{C}(t),\hat{\rho}_{tot}(0)]-\int_{0}^{t}dt'[\hat{H}_{C}(t),[\hat{H}_{J}(t'),\hat{\rho}_{tot}(t')]] -\int_{0}^{t}dt'[\hat{H}_{C}(t),[\hat{H}_{C}(t'),\hat{\rho}_{tot}(t')]].\notag\\
\end{eqnarray}
Since, the state $\rho_{E}$ of the environment (the two atomic baths) is considered as equilibrium thermal state in all the time, it is reasonable to write total density matrix of the system and environment as direct product $\rho_{tot}(t)=\rho(t)\rho_{E}$ under their weak coupling condition. A trace over all states of the environment leads to the reduced density matrix $Tr[\rho_{tot}(t)]=\rho(t)$, which induces equation of motion of the system in interaction picture,
\begin{eqnarray}\label{eq:RSIIE}
\frac{\partial}{\partial t}\hat{\rho}(t)&=&-i[\hat{H}_{J}(t),\hat{\rho}(t)]-\int_{0}^{t}dt'Tr[\hat{H}_{C}(t),[\hat{H}_{C}(t'),\hat{\rho}(t')\rho_{E}]].
\end{eqnarray}
Here, the second and the third terms in the right side of Eq. \eqref{eq:SIIE} become zero after tracing the equation over all states of the environment, since they includes single operators of the environment, $a_{\alpha k}$ or $a^{\dag}_{\alpha k}$ ($\alpha=L$, $R$). For the large number of environment degree of freedom, effective time range is within the delta function $\delta(t-t')$. In this case, weak coupling between system and environment could be treated as Markovian process~\cite{Scully}. It allow us to replace $t'$ with $t$ in $\hat{\rho}(t')$, $e^{-i\chi_{1}(t')}$ and $e^{-i\chi_{2}(t')}$ in Eq. \eqref{eq:RSIIE}, supposing change of them are much slower than the delta function.

Additionally, one needs four more steps to obtain quantum master equation for the system evolution. The four steps are~\cite{Lai12}: (a) taking the trace; (b) carrying out the time integral; (c) summing over all states of the electrodes through the equivalent relation between $\sum_{k}$ and $\int d\epsilon_{\alpha k} D(\epsilon_{\alpha k})$, where $D(\epsilon_{\alpha k})$ is density of energy state in the atomic baths; (d) using the system free evolution Hamiltonian $H_{0}=\sum_{l=1}^{2}\varepsilon_{l}n_{l}$, one can transforms Eq. \eqref{eq:RSIIE} into Schrodinger picture. In this way, one achieves a variable coefficient quantum master equation,
\begin{eqnarray}\label{eq:QME}
\frac{\partial}{\partial t}\rho&=&-i[\sum_{l=1}^{2}\varepsilon_{l}n_{l}-J(e^{i\phi(t)}a_{1}^{\dag}a_{2}+H.c.),\rho]+\mathcal{L}_{L}\rho+\mathcal{L}_{R}\rho,
\end{eqnarray}
where the system density matrix is given in schr\"{o}dinger picture $\rho=e^{-itH_{0}}\hat{\rho} e^{itH_{0}}$. The first term on the right side of Eq. \eqref{eq:QME} represents coherent evolution of the center system. The Lindblad super operators reveal atom leakage between the system and the two atomic baths. They have the typical formulations as
\begin{eqnarray}\label{eq:LSOL}
\mathcal{L}_{L}\rho&=&\frac{\Gamma_{L}}{2}(1-f_{L}(\varepsilon_{1}))(2a_{1}\rho a_{1}^{\dag}-\{a_{1}^{\dag}a_{1},\rho\})+\frac{\Gamma_{L}}{2}f_{L}(\varepsilon_{1})(2a_{1}^{\dag}\rho a_{1}-\{a_{1}a_{1}^{\dag},\rho\})\end{eqnarray}
and
\begin{eqnarray}\label{eq:LSOR}
\mathcal{L}_{R}\rho&=&\frac{\Gamma_{R}}{2}(1-f_{R}(\varepsilon_{2}))(2a_{2}\rho a_{2}^{\dag}-\{a_{2}^{\dag}a_{2},\rho\})+\frac{\Gamma_{R}}{2}f_{R}(\varepsilon_{2})(2a_{2}^{\dag}\rho a_{2}-\{a_{2}a_{2}^{\dag},\rho\}). \end{eqnarray}
The braces here indicate anti commutations. The coupling rates read $\Gamma_{L}=2\pi D(\varepsilon_{1})|t_{L}|^{2}$ and $\Gamma_{R}=2\pi D(\varepsilon_{2})|t_{R}|^{2}$ with energy density $D(\epsilon_{\alpha k})$ in atomic baths. In addition, $f_{L}(\varepsilon_{1})$ and $f_{R}(\varepsilon_{2})$ are the Fermi-Dirac distribution functions $f_{L}(\varepsilon_{1})=\frac{1}{e^{(\varepsilon_{1}-\mu_{L})/k_{B}T_{L}}+1}$ and $f_{R}(\varepsilon_{2})=\frac{1}{e^{(\varepsilon_{2}-\mu_{R})/k_{B}T_{R}}+1}$, respectively. Here, $k_{B}$ is the Boltzmann constant, $T_{L}$ and $T_{R}$ represent corresponding temperature the two atomic reservoirs.

\begin{figure}
  \includegraphics[width=8.5cm]{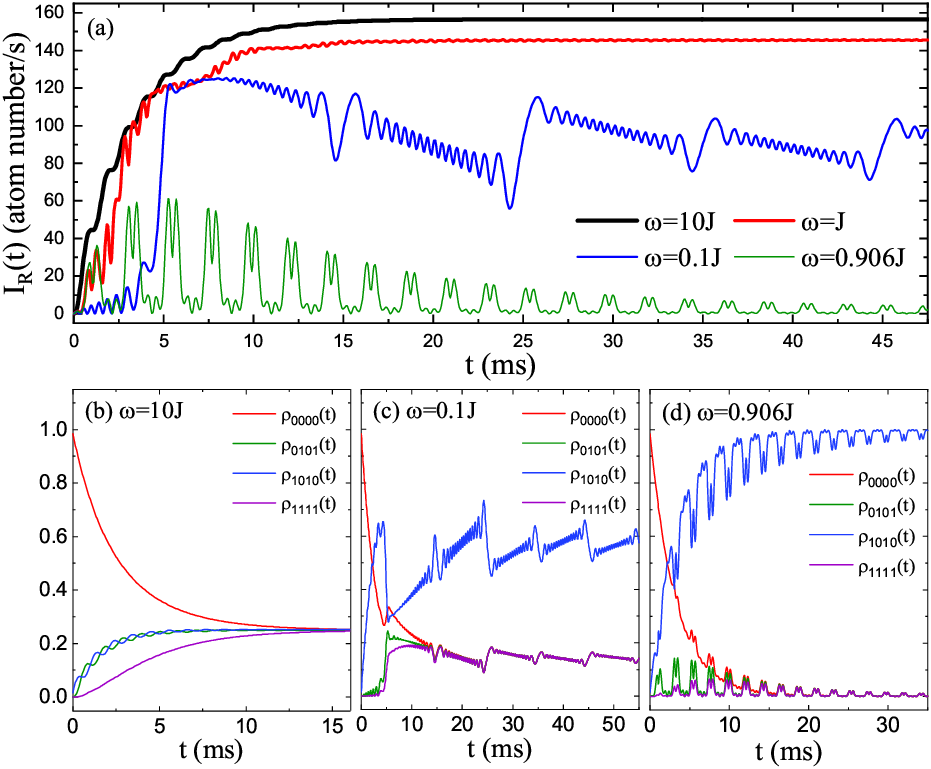}\\
  \caption{(Color on line) (a) Atomic current as a function of time with different driving frequencies at $K=5J$. (b)-(d) Time evolution of diagonal density matrix elements (probabilities of atom occupation) corresponding to the lines in (a) for different shaking frequencies.}\label{fig2}
\end{figure}

\begin{flushleft}
  \item[3.2] \textbf{Results}
\end{flushleft}
Hilbert space of the system is generated from the basic states \{$|00\rangle$, $|01\rangle$, $|10\rangle$, $|11\rangle$\}, in which $|00\rangle$ represents both energy levels $\varepsilon_{1}$ and $\varepsilon_{2}$ are empty, $|01\rangle$ denotes the level $\varepsilon_{2}$ is occupied by one atom, $|10\rangle$ represents the level $\varepsilon_{1}$ is occupied by one atom, and $|11\rangle$ represents the both levels $\varepsilon_{1}$ and $\varepsilon_{2}$ are occupied by one atom, respectively. Under these basic states, Eq.\eqref{eq:QME} can be written as
\begin{eqnarray}\label{eq:EOM-Matrix}
\frac{\partial}{\partial t}\rho_{j_{1}j_{2}k_{1}k_{2}}(t)=\sum_{l_{1},l_{2}}M_{j_{1}j_{2}l_{1}l_{2}}(t)\rho_{l_{1}l_{2}k_{1}k_{2}}(t).
\end{eqnarray}
where the matrix elements are defined as $\rho_{j_{1}j_{2}k_{1}k_{2}}(t)=\langle j_{1}j_{2}|\rho(t)|k_{1}k_{2}\rangle$, $M_{j_{1}j_{2}l_{1}l_{2}}(t)=\langle j_{1}j_{2}|M(t)|l_{1}l_{2}\rangle$, and the indices take $j_{1}$, $j_{2}$, $k_{1}$, $k_{2}$, $l_{1}$, $l_{2}=0$, $1$. In this way, Eq.\eqref{eq:EOM-Matrix} can be written as a Matrix differential equation
\begin{eqnarray}\label{eq:EOM-formal}
\frac{\partial}{\partial t}\rho(t)=M(t)\rho(t).
\end{eqnarray}
Next, to solve Eq.\eqref{eq:EOM-formal}, it could be transformed into equivalent integral equation,
\begin{eqnarray}\label{eq:EOM-integral}
\rho(t)=\rho(0)+\int_{0}^{t}M(t_{1})\rho(t_{1})dt_{1}.
\end{eqnarray}
Here, $\rho(0)$ represents initial state of the system. Equation of this type can be solved by iteration in principle. The process of successive re-insertion of the left-hand side of Eq.\eqref{eq:EOM-integral} leads to the Neumann series
\begin{eqnarray}\label{eq:SOL-Neumann}
\rho(t)=[1+\int_{0}^{t}dt_{1}M(t_{1})+\int_{0}^{t}dt_{1}\int_{0}^{t_{1}}dt_{2}M(t_{1})M(t_{2})+\int_{0}^{t}dt_{1}\int_{0}^{t_{1}}dt_{2}\int_{0}^{t_{2}}dt_{3}M(t_{1})M(t_{2})M(t_{3})+...]\rho(0).
\end{eqnarray}
Considering $M(t_{1})$ does not commute with $M(t_{2})$ for $t_{1}\neq t_{2}$, using Dyson product to Eq.\eqref{eq:SOL-Neumann}, we reach a solution of the time ordered product form
\begin{eqnarray}\label{eq:SOL-Dyson}
\rho(t)=[1+\sum_{n=1}^{\infty}\frac{1}{n!}\int_{0}^{t}dt_{1}...\int_{0}^{t}dt_{n}\mathcal{T}(M(t_{1})...M(t_{n}))]\rho(0),
\end{eqnarray}
where, $\mathcal{T}$ is the time-ordering operator. It is possible to formally sum up the series \eqref{eq:SOL-Dyson}, arriving at the time-ordered exponential function
\begin{eqnarray}\label{eq:SOL-exp}
\rho(t)=\mathcal{T}e^{\int_{0}^{t}M(t')dt'}\rho(0).
\end{eqnarray}
Eq.\eqref{eq:SOL-exp} is final solution for the equation of motion of the system\eqref{eq:EOM-formal}. Since Eq.\eqref{eq:SOL-exp} is just a formal solution, it is hard to calculate it directly. For the following numerical treatment, we use the approximate equation by removing the time-ordering operator
\begin{eqnarray}\label{eq:apx-SOL-exp}
\rho(t)\approx e^{\int_{0}^{t}M(t')dt'}\rho(0).
\end{eqnarray}
Solution Eq.\eqref{eq:apx-SOL-exp} is exact when $M(t_{1})$ commutes with $M(t_{2})$ for any $t_{1}$ and $t_{2}$. For convenience, we further note the evolution matrix as $N(t)=\int_{0}^{t}M(t')dt'$. In the instantaneous eigenstates approach, exponential function of the evolution matrix $N(t)$ would be decomposed into $e^{N(t)}=V(t)e^{\Lambda(t)}V^{-1}(t)$, where $V(t)$ and $\Lambda(t)$ are instantaneous eigenvector and eigenvalue matrices of $N(t)$ at time $t$, respectively. This process allows us to obtain the time dependent density matrix $\rho(t)$ for the system evolution.

In the differential equation \eqref{eq:EOM-Matrix}, just $6$ matrix elements would be involved, they are $\rho_{0000}$, $\rho_{0101}$, $\rho_{1010}$, $\rho_{1111}$, $\rho_{0110}$ and $\rho_{1001}$. For initial state of the system $\rho(0)$, we suppose both two wells are always empty at the beginning, namely $\rho_{0000}(0)=1$, $\rho_{0101}(0)=\rho_{1010}(0)=\rho_{1111}(0)=\rho_{0110}(0)=\rho_{1001}(0)=0$. During the following numerical calculations, the basic parameters are set to be $J=2\pi\times 500$ Hz~\cite{Livi}, $\varepsilon_{1}=\varepsilon_{2}=5J$, $\mu_{L}=10J$, $\mu_{R}=0$ and $k_{B}T_{L}=k_{B}T_{R}=0.01J$ throughout this paper. In addition, the system-reservoir coupling rate is taken to be $\Gamma_{L}=\Gamma_{R}=0.1J$ except particularly indicated.

As outcomes, atomic currents are derived from the continuity equation $I_{L}-I_{R}=\frac{\partial}{\partial t}Tr_{S}[\rho(t)\sum_{l=1}^{2}n_{l}]$~\cite{Davies,Jauho,Twamley}, where $Tr_{S}$ indicates trace over all the system states. Current $I_{R}$ on the right side of the system, which is equivalent to the left side current $I_{L}$ according to Kirchhoff's circuit laws,  has the detail expression,
\begin{eqnarray}\label{eq:CC}
    I_{R}&=&-\Gamma_{R}f_{R}(\varepsilon_{2})(\rho_{0000}+\rho_{1010})+\Gamma_{R}(1-f_{R}(\varepsilon_{2}))(\rho_{0101}+\rho_{1111}).
\end{eqnarray}
where $\rho_{0000}$, $\rho_{0101}$, $\rho_{1010}$ and $\rho_{1111}$ are diagonal elements of the density matrix $\rho$ given in Eq. \eqref{eq:EOM-Matrix}.

\begin{figure}
  \includegraphics[width=8.5cm]{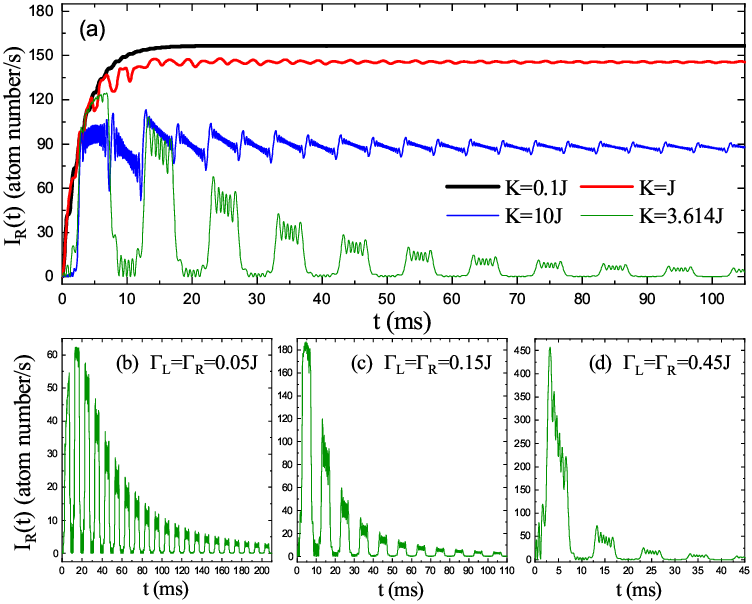}\\
  \caption{(Color on line) (a) Atomic current as a function of time with different driving amplitudes in the case of $\omega=0.2J$. In (b)-(d), the destructed current at the driving amplitude $K=3.614J$ in (a) has is plotted under different system-reservoir coupling rates.}\label{fig3}
\end{figure}

The atomtronic transistor displays stationary current under much high driving frequency $\omega\gg J$ at a certain driving amplitude $K$ as illustrated in Fig.~\ref{fig2} (a). For lower driving frequency or larger driving amplitude, current would fluctuate. Obviously, shaking induced current decrease is demonstrated here, which is consistent with coherent tunneling destruction in optical lattices reported previously~\cite{Creffield6,Lignier,Kierig,Eckardt,Zenesini}. Especially, in Fig.~\ref{fig2}, currents fluctuate a short time and then tend to disappear near the chosen parameters $(\omega, K)=(0.906J, 5J)$. Normally, atom wave function occupies both the optical wells as shown in Figs.~\ref{fig2} (b)-(d). However, the figure also illustrates that occupation of the atom wave function could be coherently controlled by tuning the driving frequency or the driving amplitude. Especially, under the particular chosen parameters $(\omega$, $K)$, the atom wave packet would be trapped in one of the optical wells. It is similar to the coherent population trapping (CPT) in quantum optics~\cite{Whitley,Alzetta,Arimondo}, in which atoms are coherently localized in internal electronic states under driving of laser fields. The effect of coherently localized atom wave function is exactly the reason of insulation appeared above. In fact, this is the CDT in the open system which gives rise to Mott insulator in optical lattices~\cite{Eckardt,Zenesini,Creffield6}.

Destruction of current can also be controlled by tuning driving amplitude $K$. Such example is shown in Fig.~\ref{fig3} (a). It is not surprising as the Peierls phase $\phi(t)$ depends on $K$. Time length of the destruction is related to system-reservoir coupling strength of the open system. Indeed, Figs.~\ref{fig3} (b)-(d) reveal current destructs more fast under stronger system-reservoir coupling. When CPT of atom wave packet occurs, atom population in the right well (see Figs.~\ref{fig1}) escapes into the right reservoir. Therefore, current destructions have current tails as shown in Fig.~\ref{fig2} and Fig.~\ref{fig3}. Furthermore, the destruction time becomes long when the system-reservoir coupling is weak.

\begin{figure}
  \includegraphics[width=8.5cm]{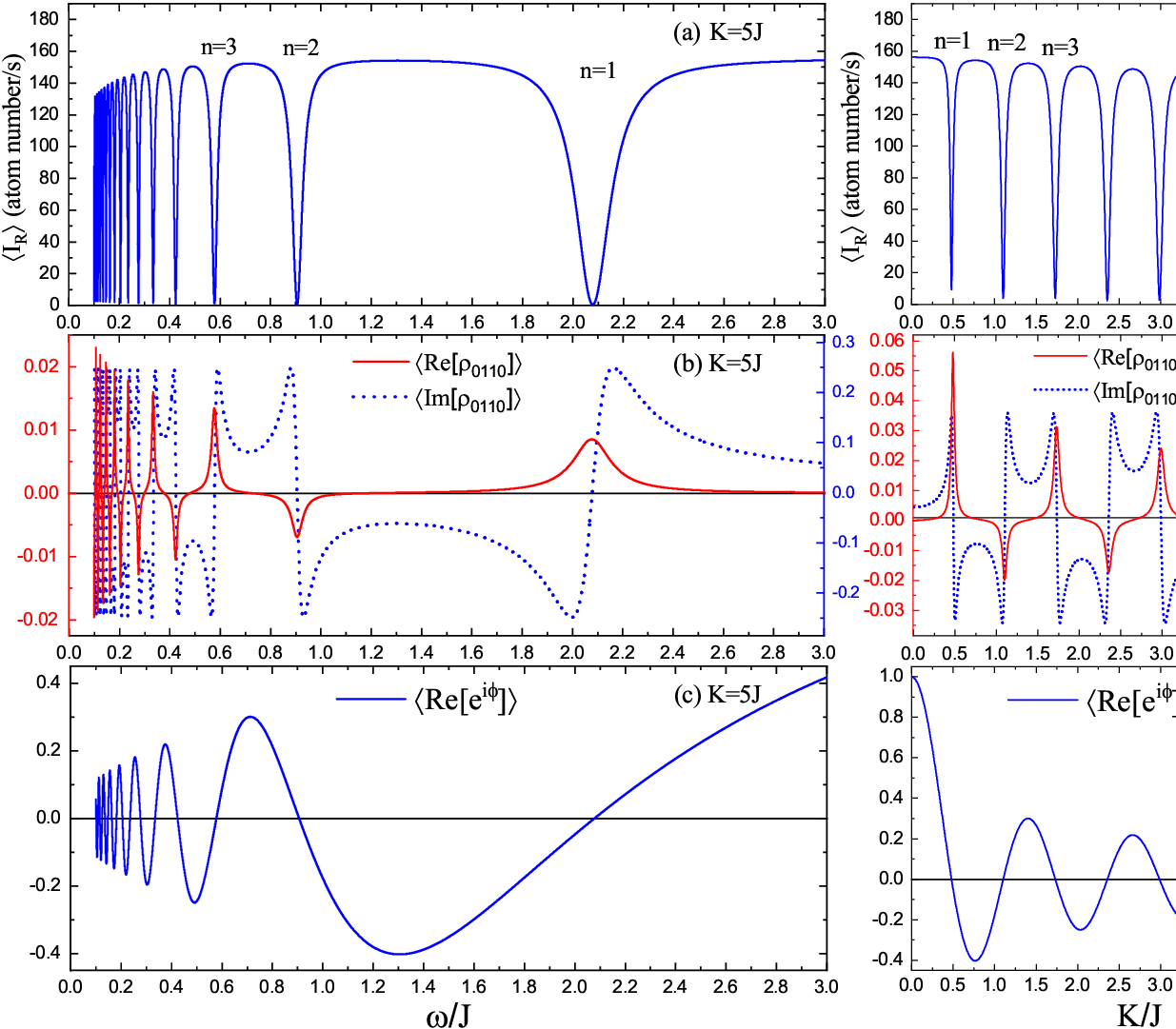}\\
  \caption{(Color on line) (a)-(c) Spectrum of the averaged current $\langle I_{R}\rangle$, the off-diagonal density matrix element $\langle \rho_{0110}\rangle$ and the phase term $\langle e^{i\phi}\rangle$ versus the shaking frequency $\omega$. (d)-(f) Spectrum of the averaged current $\langle I_{R}\rangle$, the off-diagonal density matrix element $\langle \rho_{0110}\rangle$ and the phase term $\langle e^{i\phi}\rangle$ as a function of the shaking amplitude $K$. Here, $\omega\geq 0.1J$. A black real horizontal guide line at zero point of vertical axis is drawn in each figure.}\label{fig4}
\end{figure}

Fig.~\ref{fig2} and Fig.~\ref{fig3} show currents are always positive. Therefore, let us use the formula $\langle O\rangle=\frac{1}{\tau}\int_{0}^{\tau}O(t)dt$ to calculate time averaged current and other quantities of the system. $O$ is any operator here. As currents are non-periodic, the integral range of time is taken to be $\tau=1$s which should be much longer than the period $2\pi/\omega$. Current spectrums in Fig.~\ref{fig4} (a) and (d) show many areas of insulation (the positions of current decrease) with respect to driving frequency $\omega$ and driving amplitude $K$. Since the Peierls phase $\phi(t)$ is inversely proportional to the shaking frequency $\omega$, the current valleys become very dense when the frequency tends to zero. For the convenience of discussions, these areas are signed with serial numbers $n=1$, $2$, $3$, ..., corresponding to the parameters $(\omega_{1}$, $K_{1})$, $(\omega_{2}$, $K_{2})$, $(\omega_{3}$, $K_{3})$, ..., and so on. The insulator occurs when the averaged real part of the off-diagonal density matrix element satisfies $\langle$Re$[\rho_{0110}]\rangle\neq0$ and the imaginary part satisfies $\langle$Im$[\rho_{0110}]\rangle=0$ as plotted in Fig.~\ref{fig4} (b) and (e). Therefore, the insulator effect is originated from coherence of the system. It is proved further in Fig.~\ref{fig4} (c) and (f) with the positions where $\langle$Re$[e^{i\phi}]\rangle=0$. The results demonstrate the coherence induced insulator in the driving frequency regime of both $\omega\leq J$ and $\omega>J$.

\begin{figure}
  \includegraphics[width=8.5cm]{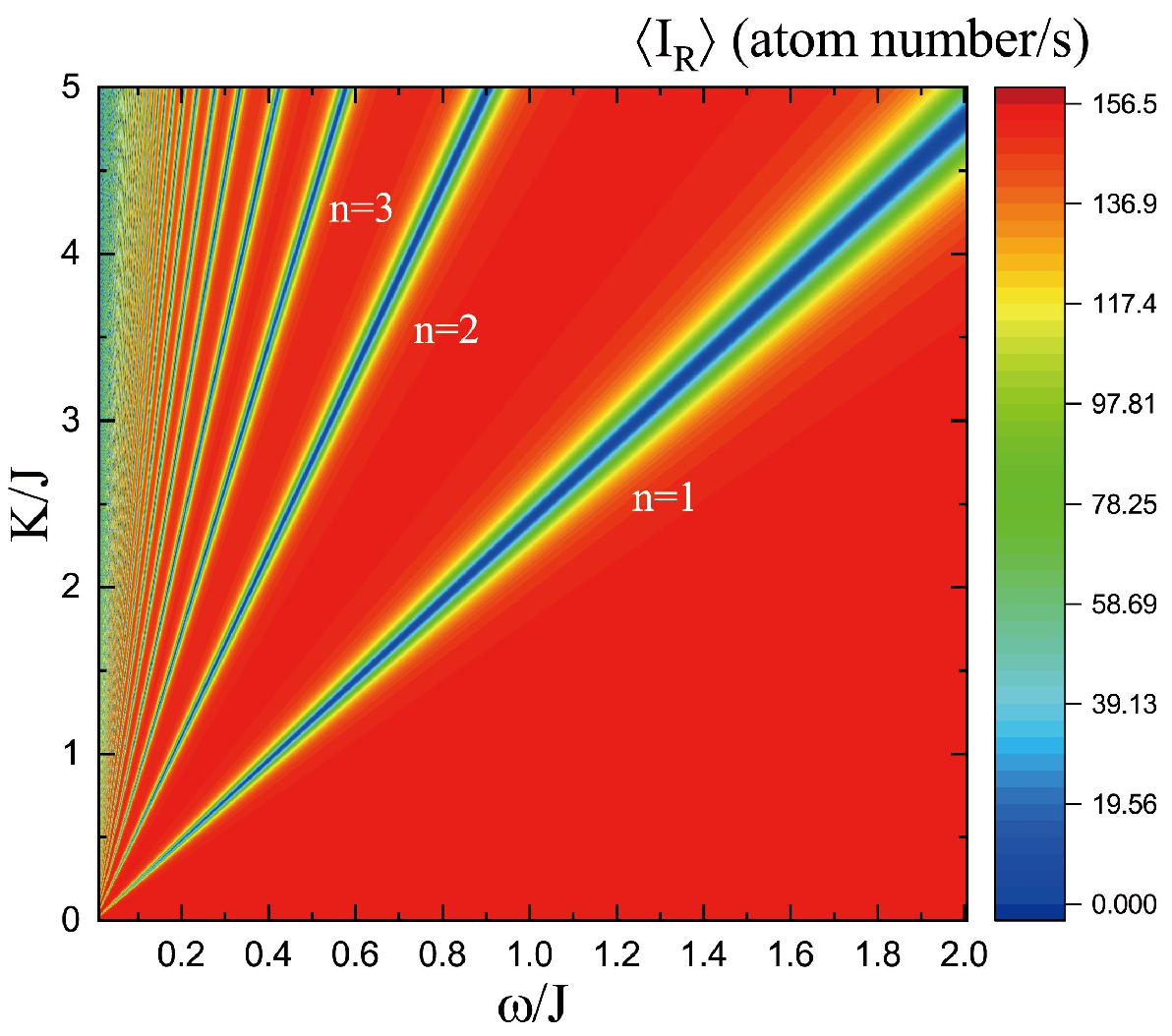}\\
  \caption{(Color on line) Current spectrum in the space of shaking frequency $\omega$ and shaking strength $K$. Blue lines are the area in which system would behaves as insulation.}\label{fig5}
\end{figure}

Current spectrum in Fig.~\ref{fig5} shows two different areas of atomic current in the space of parameters $\omega$ and $K$. In the red area, system generates nonzero current. In contrast, the blue area indicates the condition of insulator. The insulator lines in Fig.~\ref{fig5} are straight lines with different gradients calculated by $K/\omega$. Some of these gradients are arranged in Table I. Previous researches reported the ratio $K/\omega\approx2.4$ as one condition of insulator and CDT~\cite{Lignier,Eckardt,Zenesini,Luo}. This reported ratio belongs to the first ($n=1$) insulator line as shown in Table I, being one particular case of our results. It is interesting that the gradients of insulator lines $K_{1}/\omega_{1}$, $K_{2}/\omega_{2}$, $K_{3}/\omega_{3}$, ... in Fig.~\ref{fig5} approximately form an array of arithmetic sequence. Furthermore, the common difference of the sequence is nearly equal to $\pi$, namely,
\begin{eqnarray}\label{eq:AS}
\frac{K_{n+1}}{\omega_{n+1}}-\frac{K_{n}}{\omega_{n}}\approx\pi,
\end{eqnarray}
where $n$ is positive integer appeared in Fig.~\ref{fig4}, Fig.~\ref{fig5} and Table I. For $n=1$, deviation of Eq. \eqref{eq:AS} is remarkable. For $n\geq2$, it becomes more accurate. This is an empirical formula extracted from the present numerical results. Now, it becomes a common condition for realization of the coherence induced insulator in the driven optical lattice. Origin of this formula could be understood from mathematical structure of the Peierls phase $e^{i\phi}$, since Fig.~\ref{fig4} shows the series $K_{1}/\omega_{1}$, $K_{2}/\omega_{2}$, $K_{3}/\omega_{3}$, ... correspond to $\langle$Re$[e^{i\phi}]\rangle=0$ and Re$[e^{i\phi}]=\cos[\frac{K}{\omega}\sin(\omega t)]$. Actually, it is property of the zero order Bessel function $J_{0}(x)$,
\begin{eqnarray}\label{eq:Bessel}
J_{0}(x)=\sum_{k=0}^{\infty}\frac{(-1)^{k}}{(k!)^{2}}(\frac{x}{2})^{2k},
\end{eqnarray}
for any variable $x$. The chosen value $K_{n}/\omega_{n}$ is exactly the position which leads to $J_{0}(K_{n}/\omega_{n})\approx0$. Indeed, numerical estimation shows that solutions of the equation $J_{0}(x)=0$ with long enough terms satisfies the formula \eqref{eq:AS}.

Formula \eqref{eq:AS} could be used to understand the phenomenon that the insulator peaks labeled $n=1$, $2$, $3$ are running right-to-left in Fig.~\ref{fig4} (a) but left-to-right in Fig.~\ref{fig4} (d). In this formula, the ratio $\frac{K_{n}}{\omega_{n}}$ is a progressive increase function for rising of the order number $n$. In the ratio $\frac{K_{n}}{\omega_{n}}$, the amplitude $K$ is numerator and frequency $\omega$ is denominator. Therefore, to reach higher order insulator peaks, one can increase the amplitude $K$, or decrease the frequency $\omega$.

\begin{table}[hh]
  \caption{The ratio $K_{n}/\omega_{n}$ for realization of the coherence induced insulator}
  \centering
  \begin{tabular}{c c c c c c c c c c c}
    \hline
    $n$ & 1 & 2 & 3 & 4 & 5 & 6 & 7 & 8 & 9 & 10 \\[0.5ex]
    \hline
    $K_{n}/\omega_{n}$ & 2.40 & 5.52 & 8.65 & 11.79 & 14.93 & 18.07 & 21.21 & 24.35 & 27.50 & 30.63 \\
    \hline
    $n$ & 11 & 12 & 13 & 14 & 15 & 16 & 17 & 18 & 19 & 20\\[0.5ex]
    \hline
    $K_{n}/\omega_{n}$ & 33.78 & 36.92 & 40.05 & 43.20 & 46.35 & 49.48 & 52.63 & 55.76 & 58.91 & 62.05 \\
    \hline
  \end{tabular}
  \label{table}
\end{table}

\begin{flushleft}
  \item[4.] \textbf{Comparison between the instantaneous eigenstates approach and time-independent effective Hamiltonian approach}
\end{flushleft}
In the end, we would like to compare the time-dependent Hamiltonian numerical approach with the time-independent effective Hamiltonian approach. Since the original Hamiltonian \eqref{eq:OH} is time periodic, $\mathcal{H}(t)=\mathcal{H}(t+T)$, it can be expanded as
\begin{eqnarray}\label{eq:FT}
\mathcal{H}(t)&=& \sum_{m=-\infty}^{\infty}e^{im\omega t}\mathcal{H}_{m}
\end{eqnarray}
where, in contrary, $\mathcal{H}_{m}=\frac{1}{T}\int_{0}^{T}e^{-im\omega t}\mathcal{H}(t)dt$. According to Floquet engineering~\cite{Goldman,Eckardt2017,Itin2015}, typically, there is a time periodic unitary operator $U_{F}(t)=U_{F}(t+T)$ which could transform the original Hamiltonian $\mathcal{H}(t)$ into a time-independent Hamiltonian,
\begin{eqnarray}\label{eq:TIH}
H_{F}=U^{\dag}_{F}(t)\mathcal{H}(t)U_{F}(t)-i U^{\dag}_{F}(t)\frac{\partial}{\partial t}U_{F}(t).
\end{eqnarray}
Then, state of the atomtronic transistor could be written by $\rho_{tot}^{F}(t)=U^{\dag}_{F}(t)\varrho_{tot}(t)U_{F}(t)$ which satisfies the equation of motion,
\begin{eqnarray}\label{eq:EOM}
\frac{\partial}{\partial t}\rho_{tot}^{F}(t) = -i[H_{F},\rho_{tot}^{F}(t)].
\end{eqnarray}
Using the time-independent Hamiltonian $H_{F}$, one can display eigenenergy and eigenstates of the system. Therefore, the key problem in Floquet engineering is to compute the effective Hamiltonian $H_{F}$. A systematic way to calculate the effective Hamiltonian is formally given from the perturbative expansion~\cite{Goldman,Eckardt2017},
\begin{eqnarray}\label{eq:EH}
H_{F}= \sum_{\mu=0}^{\infty}H_{F}^{\mu}.
\end{eqnarray}
During actual numerical treatment, the effective Hamiltonian could be approximated by the $\mu_{cut}$th order terms, $H_{F}\approx \sum_{\mu=1}^{\mu_{cut}}H_{F}^{\mu}$, in which the first term usually takes $H_{F}^{0}=0$~\cite{Eckardt2015}. In this expansion, the driving energy $\omega$ is required to be large compared to the matrix elements of the Hamiltonian. It can be seen from the detail perturbation expansion in power of $1/\omega$,
\begin{eqnarray}\label{eq:PE}
H_{F}=H_{0}+\sum_{m=1}^{\infty}\frac{1}{m\omega}[H_{m},H_{-m}]+\sum_{m=1}^{\infty}\frac{1}{2m^{2}\omega^{2}}([[H_{m},H_{0}],H_{-m}]+[[H_{-m},H_{0}],H_{m}])+\mathcal{O}(T^{3}).
\end{eqnarray}
In this way, comparing Eq.\eqref{eq:EH} and Eq.\eqref{eq:PE}, corresponding terms in them has following relation, $H_{F}^{1}=H_{0}$, $H_{F}^{2}=\sum_{m=1}^{\infty}\frac{1}{m\omega}[H_{m},H_{-m}]$, $H_{F}^{3}=\sum_{m=1}^{\infty}\frac{1}{2m^{2}\omega^{2}}([[H_{m},H_{0}],H_{-m}]+[[H_{-m},H_{0}],H_{m}])$, ..., and so on.

\begin{figure}
  \includegraphics[width=8.5cm]{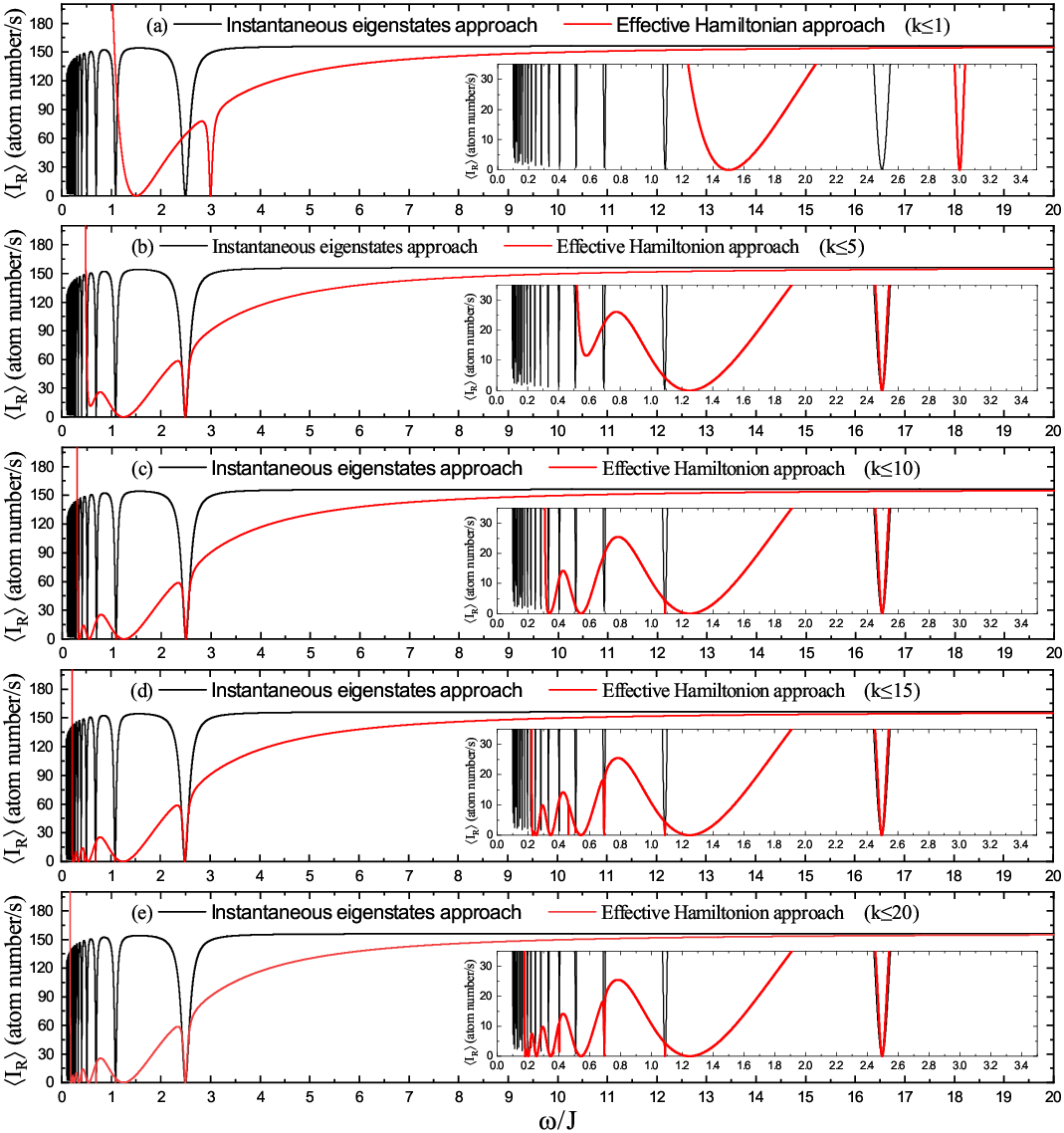}\\
  \caption{(Color on line) Comparison of results calculated from instantaneous eigenstates approach (black lines) and time-independent effective Hamiltonian approach (red lines). For the later method, (a) $2$-order expansion ($k\leq 1$), (b) $6$-order expansion ($k\leq 5$), (c) $11$-order expansion ($k\leq 10$), (d) $16$-order expansion ($k\leq 15$) and (e) $21$-order expansion ($k\leq 20$) are considered, respectively. Here, the driving amplitude is taken to be $K=6J$. The insets are enlarged figures of corresponding original lines in low frequency regime.}\label{fig6}
\end{figure}

In Eq.\eqref{eq:FT}, using the original Hamiltonian \eqref{eq:OH}, one can compute $H_{m}$ for all $m$. Then, these terms are $H_{0}=\sum_{l=1}^{2}\varepsilon_{l}n_{l}-J (a_{1}^{\dag}a_{2}+H.c.)+\sum_{\alpha,k}\epsilon_{\alpha k}n_{\alpha k}-\sum_{k}(t_{L}a_{1}^{\dag}a_{L k}+t_{R}a_{2}^{\dag}a_{R k}+H.c.)$, $H_{1}=H_{-1}=-\frac{K}{4}(n_{2}-n_{1})$, and $H_{m}=0$ ($m\geq2$). They allow us to rewrite leading orders of the iterative expansion in Eq.\eqref{eq:EH} and Eq.\eqref{eq:PE} as
\begin{eqnarray}\label{eq:FH1}
H_{F}^{1}&=&H_{0}=\sum_{l=1}^{2}\varepsilon_{l}n_{l}-J (a_{1}^{\dag}a_{2}+H.c.)+\sum_{\alpha,k}\epsilon_{\alpha k}n_{\alpha k}-\sum_{k}(t_{L}a_{1}^{\dag}a_{L k}+t_{R}a_{2}^{\dag}a_{R k}+H.c.).
\end{eqnarray}
\begin{eqnarray}\label{eq:FH2}
H_{F}^{2}&=&\frac{1}{\omega}[H_{1},H_{-1}]=0.
\end{eqnarray}
\begin{eqnarray}\label{eq:FH3}
&&H_{F}^{3}=\frac{1}{2\omega^{2}}([[H_{1},H_{0}],H_{-1}]+[[H_{-1},H_{0}],H_{1}])=(\frac{K}{4\omega})^{2}[4J (a_{1}^{\dag}a_{2}+H.c.)+\sum_{k}(t_{L}a_{1}^{\dag}a_{L k}+t_{R}a_{2}^{\dag}a_{R k}+H.c.)].
\end{eqnarray}
Consequently, Eq.\eqref{eq:EH} could be written as
\begin{eqnarray}\label{eq:EHE}
&&H_{F}\approx H_{F}^{1}+H_{F}^{2}+H_{F}^{3}+...=\sum_{l=1}^{2}\varepsilon_{l}n_{l}-J(1-4(\frac{K}{4\omega})^{2}+...)(a_{1}^{\dag}a_{2}+H.c.)\notag\\
&&+\sum_{\alpha,k}\epsilon_{\alpha k}n_{\alpha k}-\sum_{k}(1-(\frac{K}{4\omega})^{2}+...)(t_{L}a_{1}^{\dag}a_{L k}+t_{R}a_{2}^{\dag}a_{R k}+H.c.).
\end{eqnarray}

In fact, effective Hamiltonian of a harmonically driven system could be achieved exactly as emphasized in Ref.~\cite{Goldman}. Indeed, in section three, we have a unitary operator $U(t)$ with detail expression to obtain an exact time-dependent Floquet Hamiltonian \eqref{eq:TDFH}. Through taking an average within a time period $T$ to the Hamiltonian $H(t)$ given in Eq.\eqref{eq:TDFH}, $H_{eff}=\frac{1}{T}\int_{0}^{T}H(t')dt'$, the exact effective Hamiltonian could be derived,
\begin{eqnarray}\label{eq:EEH}
H_{eff}=\sum_{l=1}^{2}\varepsilon_{l}n_{l}-J_{eff} (a_{1}^{\dag}a_{2}+H.c.)+\sum_{\alpha,k}\epsilon_{\alpha k}n_{\alpha k}-\sum_{k}(J_{0}(W_{1})t_{L}a_{1}^{\dag}a_{L k}+J_{0}(W_{2})t_{R}a_{2}^{\dag}a_{R k}+H.c.),
\end{eqnarray}
where $J_{eff}=JJ_{0}(W_{2}-W_{1})$ and $W_{2}=-W_{1}=\frac{K}{2\omega}$. The zero order Bessel function $J_{0}$ appeared in this Hamiltonian is given in Eq.\eqref{eq:Bessel}. With different parameter structures, we have following expansions with infinite terms,
namely,
\begin{eqnarray}\label{eq:ZOBF1}
J_{0}(W_{2}-W_{1})=1-4(\frac{K}{4\omega})^{2}+4(\frac{K}{4\omega})^{4}+\frac{16}{9}(\frac{K}{4\omega})^{6}+...
\end{eqnarray}
and
\begin{eqnarray}\label{eq:ZOBF2}
J_{0}(W_{1})=J_{0}(W_{2})=1-(\frac{K}{4\omega})^{2}+\frac{1}{4}(\frac{K}{4\omega})^{4}+\frac{1}{36}(\frac{K}{4\omega})^{6}+...
\end{eqnarray}
It is obvious that effective Hamiltonian \eqref{eq:EEH} is absolutely in accord with systematic expansion in Eq.\eqref{eq:EHE}. Above comparison indicates that effective Hamiltonian $H_{eff}$ and $H_{F}$ are equivalent for our system. Since $H_{eff}$ has been written exactly with all terms in Eq.\eqref{eq:EEH}, we derive the following quantum mater equation based on this Hamiltonian,
\begin{eqnarray}\label{eq:EQME}
\frac{\partial}{\partial t}\rho_{eff} &=&-i[\sum_{l=1}^{2}\varepsilon_{l}n_{l}- J_{eff}(a_{1}^{\dag}a_{2}+a_{2}^{\dag}a_{1}),\rho_{eff}]+J_{0}^{2}(W_{1})\mathcal{L}_{L}\rho_{eff}+J_{0}^{2}(W_{2})\mathcal{L}_{R}\rho_{eff},
\end{eqnarray}
where $\rho_{eff}$ represents density matrix for atomic states with respect to the effective Hamiltonian. The super operators $\mathcal{L}_{L}\rho_{eff}$ and $\mathcal{L}_{R}\rho_{eff}$ are given in Eqs.\eqref{eq:LSOL} and \eqref{eq:LSOR}.

Fig.~\ref{fig6} shows that calculations from the time-independent effective Hamiltonian approach match the results of the instantaneous eigenstates approach in the regime of high frequency, where $\omega>10J$. In the low frequency region $\omega<10J$, difference between the two methods becomes very large. Obviously, the time-independent effective Hamiltonian approach depends on number of terms in expansions given in Eq.\eqref{eq:EH} or Eq.\eqref{eq:EEH}. The first two leading orders ($k\leq 1$) of this expansion just reflect one insulator point and its position in the frequency is a bit deviate from the precise value calculated from the instantaneous eigenstates approach. We call it a precise value, as this point satisfies $K/\omega=2.40$ which has been proved both experimentally~\cite{Lignier,Zenesini} and numerically~\cite{Eckardt}. When more terms in the effective Hamiltonian expansion are considered, such as $k\leq5$, $k\leq10$, $k\leq15$, and $k\leq20$, more insulator point appears in the corresponding current lines. In the lower frequency range, although current lines calculated from the two methods are different, they completely meet at these insulators points as long as enough expansion terms are considered in the effective Hamiltonian. From the point of view of insulator areas, time-independent effective Hamiltonian approach is applicable for both $\omega>J$ and $\omega<J$ regimes as illustrated in Fig.~\ref{fig6}. However, for much low driving frequency, $\omega\rightarrow 0$, the time-independent effective Hamiltonian approach displays divergence. It is originated from the perturbation expansion with power of $1/\omega$ in the effective Hamiltonian~\cite{Goldman,Eckardt2017}. In contrast, as a more general protocol, the instantaneous eigenstates approach always offers limited currents.

\begin{flushleft}
  \item[5.] \textbf{Discussions on feasibility}
\end{flushleft}

Challenges for realization of the configuration considered here are mainly originated from the construction of optical traps controlling a few atoms, to create shaking potential, and to monitor atoms. Fortunately, transport and trapping of atoms at the level of single atom could be manipulated using optical tweezers as reported recently~\cite{MEKim,Stuart}. Open systems of ultracold atoms have been implemented in experiments for atomtronic technology~\cite{Wright,Caliga17,Krinner,Amico}. Confining and tunneling of a few atoms in optical wells have been observed earlier~\cite{Folling,Cheinet}. In optical lattices, periodically vibrating potential could be produced by laser fields reflecting on a mirror which is connected to a piezo-electric actuator~\cite{Zenesini}. Atomic current could be measured by detecting atom population with single atom fluorescence counts in a particular well~\cite{Schlagheck} or detecting atom velocity with optical method such as Raman process~\cite{CWZhang}. These works imply construction of our system is available in laboratory.

\begin{flushleft}
  \item[6.] \textbf{Conclusions}
\end{flushleft}
Above analysis through the change of matrix elements in density matrix of quantum master equation reveals that phase change from conductor to insulator would be happened when the atom wave packet is coherently localized in one of the optical wells. We call it coherent localization, because, when one of the two wells is fully occupied and the other is almost empty, non-diagonal elements of the density matrix are not zero. Therefore, coherent trapping of atom wave packet plays important role for the shaking induced Mott insulator. The most significant result in this work is the empirical formula $K_{n+1}/\omega_{n+1}-K_{n}/\omega_{n}\approx\pi$ of driving amplitude and frequency for the generation of driving induced Mott insulator. It is achieved from numerical results using the instantaneous eigenstates approach which is a more general method compared with the time-independent effective Hamiltonian approach as proved here, being applicable for any driving frequencies of the driven system.

\begin{acknowledgments}
This work was supported by National Natural Science Foundation of China (Grant No. 12304190), Natural Science Foundation of Beijing Municipality (Grant No. 1252018), R \& D Program of Beijing Municipal Commission of Education (Grant No. KM202011232017), and Natural Science Foundation of Beijing Municipality (Grant No. 1232026).
\end{acknowledgments}


\begin{thebibliography}{99}

\bibitem{Seaman} Seaman B T, Kr\"{a}mer M, Anderson D Z and Holland M J 2007 Atomtronics: Ultracold-atom analogs of electronic devices Phys. Rev. A \textbf{75} 023615

\bibitem{Pepino} Pepino R A, Cooper J, Anderson D Z and Holland M J 2009 Atomtronic Circuits of Diodes and Transistors Phys. Rev. Lett. \textbf{103} 140405

\bibitem{Caliga} Caliga S C, Straatsma C J E, Zozulya A A and Anderson D Z 2016 Principles of an atomtronic transistor New J. Phys. \textbf{18} 015012

\bibitem{Caliga17} Caliga S C, Straatsma C J E and Anderson D Z 2017 Experimental demonstration of an atomtronic battery New J. Phys. \textbf{19} 013036

\bibitem{Lai} Lai W, Ma Y-Q, Zhuang L and Liu W M 2019 Photovoltaic Effect of Atomtronics Induced by an Artificial Gauge Field Phys. Rev. Lett. \textbf{122} 223202

\bibitem{JGLee} Lee J G, McIlvain B J, Lobb C J and Hill W T 2013 Analogs of Basic Electronic Circuit Elements in a Free-Space Atom Chip Sci. Rep. \textbf{3} 1034

\bibitem{Wilsmann} Wilsmann K W, Ymai L H, Tonel A P, Links J and Foerster A 2018 Control of tunneling in an atomtronic switching device Commun. Phys. \textbf{1} 91

\bibitem{Grossmann} Grossmann F, Dittrich T, Jung P and H\"{a}nggi P 1991 Coherent destruction of tunneling Phys. Rev. Lett. \textbf{67} 516

\bibitem{Grifoni} Grifoni M and H\"{a}nggi P 1998 Driven quantum tunneling Phys. Rep. \textbf{304} 229

\bibitem{Kierig} Kierig E, Schnorrberger U, Schietinger A, Tomkovic J and Oberthaler M K 2008 Single-Particle Tunneling in Strongly Driven Double-Well Potentials Phys. Rev. Lett. \textbf{100} 190405

\bibitem{Lignier} Lignier H, Sias C, Ciampini D, Singh Y, Zenesini A, Morsch O and Arimondo E. Dynamical Control of Matter-Wave Tunneling in Periodic Potentials Phys. Rev. Lett. 2007 \textbf{99} 220403

\bibitem{Zenesini} Zenesini A, Lignier H, Ciampini D, Morsch O and Arimondo E 2009 Coherent Control of Dressed Matter Waves Phys. Rev. Lett. \textbf{102} 100403

\bibitem{Eckardt} Eckardt A, Weiss C and Holthaus M 2005 Superfluid-Insulator Transition in a Periodically Driven Optical Lattice Phys. Rev. Lett. \textbf{95} 260404.

\bibitem{Creffield6} Creffield C E and Monteiro T S 2006 Tuning the Mott Transition in a Bose-Einstein Condensate by Multiple Photon Absorption Phys. Rev. Lett. \textbf{96} 210403

\bibitem{Liberto} Di Liberto M, Comparin T, Kock T, \"{O}lschl\"{a}ger M, Hemmerich A and Smith C M 2014 Controlling coherence via tuning of the population imbalance in a bipartite optical lattice Nat. Commun. \textbf{5} 5735

\bibitem{Sorensen} S{\o}rensen A S, Demler E and Lukin M D 2005 Fractional Quantum Hall States of Atoms in Optical Lattices Phys. Rev. Lett. \textbf{94} 086803

\bibitem{Miao} Miao S, Zhang Z, Zhao Y, Zhao Z, Wang H and Hu J 2022 Bosonic fractional quantum Hall conductance in shaken honeycomb optical lattices without flat bands Phys. Rev. B \textbf{106} 054310

\bibitem{Kitagawa} Kitagawa T, Berg E, Rudner M and Demler E 2010 Topological characterization of periodically driven quantum systems Phys. Rev. B \textbf{82} 235114

\bibitem{Rudner} Rudner M S, Lindner N H, Berg E and Levin M 2013 Anomalous Edge States and the Bulk-Edge Correspondence for Periodically Driven Two-Dimensional Systems Phys. Rev. X \textbf{3} 031005

\bibitem{Zheng} Zheng W and Zhai H 2014 Floquet topological states in shaking optical lattices Phys. Rev. A \textbf{89} 061603(R).

\bibitem{Wintersperger} Wintersperger K, Braun C, Nur\"{U}nal F, Eckardt A, Liberto M D, Goldman N, Bloch I and Aidelsburger M 2020 Realization of an anomalous Floquet topological system with ultracold atoms Nat. Phys. \textbf{16} 1058

\bibitem{Cheng} Cheng S, Yin H, Lu Z, He C, Wang P and Xianlong G 2020 Predicting Large-Chern-Number Phases in a Shaken Optical Dice Lattice Phys. Rev. A \textbf{101} 043620

\bibitem{JYZhang} Zhang J-Y, Yi C-R, Zhang L, Jiao R-H, Shi K-Y, Yuan H, Zhang W, Liu X-J, Chen S and Pan J-W 2023 Tuning anomalous Floquet topological bands with ultracold atoms Phys. Rev. Lett. \textbf{130} 043201

\bibitem{Mei} Mei F, You J-B, Zhang D-W, Yang X C, Fazio R, Zhu S-L and Kwek L C 2014 Topological insulator and particle pumping in a one-dimensional shaken optical lattice Phys. Rev. A \textbf{90} 063638

\bibitem{Kang} Kang J H and Shin Y I 2020 Topological Floquet engineering of a one-dimensional optical lattice via resonant shaking with two harmonic frequencies Phys. Rev. A \textbf{102} 063315

\bibitem{Feng} Feng Y, Fan J, Zhou X, Chen G and Jia S 2019 Topological superradiance in a shaken dynamical optical lattices Phys. Rev. A \textbf{99} 043630

\bibitem{Song} Song B, Dutta S, Bhave S, Yu J-C, Carter E, Cooper N and Schneider U 2022 Realizing discontinuous quantum phase transitions in a strongly-correlated driven optical lattice Nat. Phys. \textbf{18} 259

\bibitem{Struck} Struck J, \"{O}lschl\"{a}ger C, Weinberg M, Hauke P, Simonet J, Eckardt A, Lewenstein M, Sengstock K and Windpassinger P 2012 Tunable Gauge Potential for Neutral and Spinless Particles in Driven Optical Lattices Phys. Rev. Lett. \textbf{108} 225304

\bibitem{Hauke} Hauke P, Tieleman O, Celi A, \"{O}lschl\"{a}ger C, Simonet J, Struck J, Weinberg M, Windpassinger P, Sengstock K, Lewenstein M and Eckardt A 2012 Non-Abelian Gauge Fields and Topological Insulators in Shaken Optical Lattices Phys. Rev. Lett. \textbf{109} 145301

\bibitem{Creffield} Creffield C E, Pieplow G, Sols F and Goldman N 2016 Realization of uniform synthetic magnetic fields by periodically shaking an optical square lattice New J. Phys. \textbf{18} 093013

\bibitem{Price} Price H M, Ozawa T and Goldman N 2017 Synthetic dimensions for cold atoms from shaking a harmonic trap Phys. Rev. A \textbf{95} 023607

\bibitem{Blanco} Blanco-Mas P and Creffield C E 2023 Generating soliton trains through Floquet engineering Phys. Rev. A \textbf{107} 043310

\bibitem{Wang} Wang K, Xiong F, Long Y, Ma Y and Parker C V 2023 Instability and momentum bifurcation of a molecular Bose-Einstein condensate in a shaken lattice with exotic dispersion Phys. Rev. A \textbf{108} L051302

\bibitem{Heinze} Heinze J, Krauser J S, Fl\"{a}schner N, Hundt B, G\"{o}tze S, Itin A P, Mathey L, Sengstock K and Becker C 2013 Intrinsic Photoconductivity of Ultracold Fermions in Optical Lattices Phys. Rev. Lett. \textbf{110} 085302

\bibitem{Maricq} Maricq M M 1982 Application of average Hamiltonian theory to the NMR of sohds Phys. Rev. B \textbf{25} 6622

\bibitem{Grozdanov} Grozdanov T P and Rakovi\'{c} M J 1988 Quantum system driven by rapidly varying periodic perturbation Phys. Rev. A \textbf{38} 1739

\bibitem{Goldman} Goldman N and Dalibard J 2014 Periodically Driven Quantum Systems: Effective Hamiltonians and Engineered Gauge Fields Phys. Rev. X \textbf{4} 031027

\bibitem{Eckardt2017} Eckardt A 2017 Colloquium: Atomic quantum gases in periodically driven optical lattices Rev. Mod. Phys. \textbf{89} 011004.

\bibitem{Sun} Sun G and Eckardt A 2020 Optimal frequency window for Floquet engineering in optical latticess Phys. Rev. Research \textbf{2} 013241.



\bibitem{Bastidas} Bastidas V M, Engelhardt G, P\'{e}rez-Fern\'{a}ndez P, Vogl M and Brandes T 2014 Critical quasienergy states in driven many-body systems Phys. Rev. A \textbf{90} 063628

\bibitem{Abanin} Abanin A D, De Roeck W, Ho W W and Huveneers F 2017 Effective Hamiltonians, prethermalization, and slow energy absorption in periodically driven many-body systems Phys. Rev. B \textbf{95} 014112

\bibitem{Yang2018} Yang X, Huang B and Wang Z 2018 Floquet Topological Superfluid and Majorana Zero Modes in Two-Dimensional Periodically Driven Fermi Systems. Sci. Rep. \textbf{8} 2243

\bibitem{Jangjan} Jangjan M and Hosseini M V 2020 Floquet engineering of topological metal states and hybridization of edge states with bulk states in dimerized two-leg ladders Sci. Rep. \textbf{10} 14256

\bibitem{Lindner} Lindner N H, Refael G and Galitski V 2011 Floquet topological insulator in semiconductor quantum wells Nat. Phys. \textbf{7} 490

\bibitem{Fleury} Fleury R, Khanikaev A B and Al\`{u} A 2016 Floquet topological insulators for sound Nat. Commun. \textbf{7} 11744


\bibitem{Rodriguez} Rodriguez-Vega M, Lentz M and Seradjeh B 2022 Controlling Floquet states on ultrashort time scales Nat. Commun. \textbf{13} 7103

\bibitem{Redondo} Redondo Y V I, Xu X, Liew T C H, Ostrovskaya E A, Stegmaier A, Thomale R, Schneider C, Dam S, Klembt S, H\"{o}fling S, Tarucha S and Fraser M D 2022 Non-reciprocal band structures in an exciton-olariton Floquet optical lattice Nat. Photonics. \textbf{13} 7103

\bibitem{Maczewsky} Maczewsky L J, Zeuner J M, Nolte S and Szameit A 2017 Observation of photonic anomalous Floquet topological insulators Nat. Commun. \textbf{8} 13756

\bibitem{Liu2024} Liu W, Liu Q, Ni X, Jia Y, Ziegler K, Al\`{u} A and Chen F 2024 Floquet parity-time symmetry in integrated photonics Nat. Commun. \textbf{15} 946

\bibitem{Itin2015} Itin A P and Katsnelson M I 2015 Effective Hamiltonians for Rapidly Driven Many-Body Lattice Systems: Induced Exchange Interactions and Density-Dependent Hoppings Phys. Rev. Lett. \textbf{115} 075301

\bibitem{Itin2008} Itin A P, Watanabe S and Konotop V V 2008 Nonlinear dynamical instabilities of a condensate system in an atom-molecule dark state Phys. Rev. A \textbf{77} 043610

\bibitem{Itin} Itin A P and Neishtadt A I 2014 Effective Hamiltonians for fastly driven tight-binding chains Phys. Lett. A \textbf{378} 822

\bibitem{Shirai} Shirai T, Mori T and Miyashita S 2015 Condition for emergence of the Floquet-Gibbs state in periodically driven open systems Phys. Rev. E \textbf{91} 030101(R)

\bibitem{Liu2015} Liu D E 2015 Classification of the Floquet statistical distribution for time-periodic open systems Phys. Rev. B \textbf{91} 144301

\bibitem{Restrepo} Restrepo S, Cerrillo J, Bastidas V M, Angelakis D G and Brandes T 2016 Driven Open Quantum Systems and Floquet Stroboscopic Dynamics Phys. Rev. Lett. \textbf{117} 250401

\bibitem{Dai} Dai C M, Shi Z C and Yi X X 2016 Floquet theorem with open systems and its applications Phys. Rev. E \textbf{93} 032121

\bibitem{Schnell} Schnell A, Wu L-N, Widera A and Eckardt A 2023 Floquet-heating-induced Bose condensation in a scarlike mode of an open driven optical-lattice system Phys. Rev. A \textbf{107} L021301

\bibitem{Qian} Qian Y, Gong M and Zhang C 2011 Quantum Transport of Bosonic Cold Atoms in Double Well Optical Lattices Phys. Rev. A \textbf{84} 013608

\bibitem{Krinner} Krinner S, Esslinger T and Brantut J-P 2017 Two-terminal transport measurements with cold atoms J. Phys.: Condens. Matter \textbf{29} 343003

\bibitem{MEKim} Kim M E, Chang T-H, Fields B M, Chen C-A and Hung C-L 2019 Trapping single atoms on a nanophotonic circuit with configurable tweezer lattices Nat. Commun. \textbf{10} 1647

\bibitem{Stuart} Stuart D and Kuhn A 2018 Single-atom trapping andtransport in DMD-controlled optical tweezers New J. Phys. \textbf{20} 023013


\bibitem{Greiner} Greiner M, Mandel O, Esslinger T, H\"{a}nsch T W and Bloch I 2002 Quantum phase transition from a superfluid to a Mott insulator in a gas of ultracold atoms Nature \textbf{415} 39

\bibitem{Gajdacz} Gajdacz M, Opatrn\'{y} T and Das K K 2014 An atomtronics transistor for quantum gates Phys. Lett. A \textbf{378} 1919

\bibitem{Cao} Cao L, Brouzos I, Chatterjee B and Schmelcher P 2012 The impact of spatial correlation on the tunneling dynamics of few-boson mixtures in a combined triple well and harmonic trap New J. Phys \textbf{14} 093011

\bibitem{Vega} Rodriguez-Vega M, Lentz M and Seradjeh B 2018 Floquet perturbation theory: formalism and application to low frequency limit New J. Phys. \textbf{20} 093022

\bibitem{Scully} Scully M O and Zubairy M S. Quantum Optics (Cambridge University Press) 1997

\bibitem{Lai12} Lai W, Cao Y and Ma Z 2012 Current-oscillator correlation and Fano factor spectrum of quantum shuttle with finite bias voltage and temperature J. Phys.: Condens. Matter \textbf{24} 175301

\bibitem{Livi} Livi L F, Cappellini G, Diem M, Franchi L, Clivati C, Frittelli M, Levi F, Calonico D, Catani J, Inguscio M and Fallani L 2016 Synthetic Dimensions and Spin-Orbit Coupling with an Optical Clock Transition Phys. Rev. Lett. \textbf{117} 220401

\bibitem{Davies} Davies J H, Hershfield S, Hyldgaard P and Wilkins J W 1993 Current and rate equation for resonant tunneling Phys. Rev. B \textbf{47} 4603

\bibitem{Jauho} Jauho A P, Wingreen N S and Meir Y 1994 Time-dependent transport in interacting and noninteracting resonant-tunneling systems Phys. Rev. B \textbf{50} 5528

\bibitem{Twamley} Twamley J, Utami D W, Goan H S and Milburn G 2006 Spin-detection in a quantum electromechanical shuttle system New J. Phys. \textbf{8} 63

\bibitem{Whitley} Whitley R M and Stroud C R 1976 Double optical resonance Phys. Rev. A \textbf{14} 1498

\bibitem{Alzetta} Alzetta G, Gozzini A, Moi L and Orriols G 1976 An experimental method for the observation of RF transitions and laser beat resonances in oriented Na vapour Nuovo Cimento Soc. Ital. Fis. B \textbf{36} 5

\bibitem{Arimondo} Arimondo E 1996 V Coherent Population Trapping in Laser Spectroscopy Prog. Opt. \textbf{35} 257

\bibitem{Luo} Luo X, Li L, You L and Wu B 2014 Coherent destruction of tunneling and dark Floquet state New J. Phys. \textbf{16} 013007

\bibitem{Eckardt2015} Eckardt A and Anisimovas E 2015 High-frequency approximation for periodically driven quantum systems froma Floquet-space perspective New J. Phys. \textbf{17} 093039

\bibitem{Wright} Wright K C, Blakestad R B, Lobb C J, Phillips W D and Campbell G K 2013 Driving Phase Slips in a Superfluid Atom Circuit with a Rotating Weak Link Phys. Rev. Lett. \textbf{110} 025302

\bibitem{Amico} Amico L, Anderson D, Boshier M, Brantut J-P, Kwek L-C, Minguzzi A and von Klitzing W 2022 Colloquium: Atomtronic circuits: From many-body physics to quantum technologies Rev. Mod. Phys. \textbf{94} 041001

\bibitem{Folling} F\"{o}lling S, Trotzky S, Cheinet P, Feld M, Saers R, Widera A, M\"{u}ller T and Bloch I 2007 Direct observation of second-order atom tunnelling Nature \textbf{448} 1029

\bibitem{Cheinet} Cheinet P, Trotzky S, Feld M, Schnorrberger U, Moreno-Cardoner M, F\"{o}lling S and Bloch I 2008 Counting Atoms Using Interaction Blockade in an Optical Superlattice Phys. Rev. Lett. \textbf{101} 090404

\bibitem{Schlagheck} Schlagheck P, Malet F, Cremon J C and Reimann S M 2010 Transport and interaction blockade of cold bosonic atoms in a triple-well potential New J. Phys. \textbf{12} 065020

\bibitem{CWZhang} Zhang C, Tewari S, Lutchyn R M and Das Sarma S 2008 px+py Superfluid from s-Wave Interactions of Fermionic Cold Atoms Phys. Rev. Lett. \textbf{101} 160401


\end{thebibliography}
\end{document}